\documentclass[aps,amssymb,amsmath,prl,graphicx,twocolumn,showpacs,superscriptadress,10pt]{revtex4-1}

\usepackage{graphics}
\usepackage[pdftex]{graphicx}
\usepackage[utf8x]{inputenc}
\usepackage{bm}
\usepackage{braket}

\newcommand{\dg}{\ensuremath{^\circ}\mathrm{C}}

\newcommand{\kHz}{\mathrm{kHz}}
\newcommand{\MHz}{\mathrm{MHz}}
\newcommand{\GHz}{\mathrm{GHz}}

\newcommand{\mm}{\mathrm{mm}}
\newcommand{\m}{\mathrm{m}}
\newcommand{\nm}{\mathrm{nm}}

\newcommand{\us}{\mu \mathrm{s}}

\newcommand{\mA}{\mathrm{mA}}

\newcommand{\uW}{\mu\mathrm{W}}
\newcommand{\mW}{\mathrm{mW}}

\newcommand{\cf}{\ensuremath{^{40}\mathrm{Ca}^+\,}}
\newcommand{\ct}{\ensuremath{^{43}\mathrm{Ca}^+\,}}
\newcommand{\up}{\ensuremath{\Ket{\uparrow}}\,}
\newcommand{\dn}{\ensuremath{\Ket{\downarrow}}\,}

\newcommand{\ish}{\mbox{$\sim$}\,}

\begin{document}

\title{Optical injection and spectral filtering of high-power UV laser diodes}
\author{V.M. Schäfer}
\email{Corresponding author: vera.schafer@physics.ox.ac.uk}
\author{C.J. Ballance}
\author{C.J. Tock}
\author{D.M. Lucas}

\affiliation{Clarendon Laboratory, Department of Physics, University of Oxford, Parks Road, Oxford OX1 3PU, UK}

\date{Compiled \today, v2arXiv}



\begin{abstract}
We demonstrate injection-locking of high-power laser diodes operating at 397nm. We achieve stable operation with injection powers of $\ish100 \uW$ and a slave laser output power of up to 110mW. 
We investigate the spectral purity of the slave laser light via photon scattering experiments on a single trapped \cf ion. We show that it is possible to achieve a scattering rate indistinguishable from that of monochromatic light by filtering the laser light with a diffraction grating to remove amplified spontaneous emission. 
\end{abstract}


\maketitle

Many experiments involving atomic ions require high-power ultraviolet c.w.\ laser light, which is in general technically difficult to produce. The most common method for achieving high powers in the UV is to use cavity-enhanced second harmonic generation to frequency-double visible laser light. Typical visible light sources employed are dye lasers \cite{DyeWin}, sum-frequency mixed fibre lasers \citep{LasWIL, LasHLO}, or amplified diode lasers \cite{InjNML}. Violet/UV solid-state diode lasers are an attractive alternative to these technically complex sources, and are presently available in wavelengths close to the resonant transitions of $\mathrm{Ca}^+$, $\mathrm{Sr}^+$, $\mathrm{Ba}^+$ and $\mathrm{Yb}^+$ ions. Diode lasers typically possess very low intensity noise, but inferior spectral purity compared with frequency-doubled sources. 

Trapped atomic ions have important applications in atomic clocks \cite{ClkDid}, precision spectroscopy \cite{SpecSCHMIDT, SpecROOS}, and quantum computing \cite{ScalMON}. The main challenges for quantum computing are achieving sufficiently high-fidelity logic operations, and scaling up the number of qubits to the large numbers required for useful information processing. For trapped-ion quantum computing, high-power solid-state lasers would assist in addressing both these challenges. The highest-fidelity two-qubit quantum logic gate so far demonstrated was implemented using a Raman laser transition in \ct at $397\nm$, and required $\approx 10\mW$ of c.w. laser power at the ions \cite{HftqCJB}. Due to losses from acousto-optic modulators, optic fibres, and intensity stabilization systems, the power required at the laser source was $\approx 100\mW$. According to our error model, the dominant contribution to the gate error was due to photon scattering; this can be reduced if more laser power is available \cite{ErrOIB}. 
The Raman scattering error-per-gate for fixed gate time $t_g$ scales with $P_\mathrm{Raman}\propto 1/\mathcal{P}$, where $\mathcal{P}$ is the Raman laser power. Since $t_g\propto 1/\Omega_R$, where $\Omega_R$ is the effective Rabi frequency, and  $\Omega_R \propto \ \mathcal{P}/\Delta$ this means that for higher laser power the detuning $\Delta$ from resonance of the $4S_{1/2}-4P_{1/2}$ $397\nm$ transition can be increased, while maintaining fixed $t_g$. Since the Raman scattering rate scales with $\Gamma\propto \mathcal{P}/ \Delta^2$ fewer photons are scattered and the gate error decreases with increasing laser power. 
If one is interested in making a useful quantum computer it is necessary to significantly scale up the laser systems; due to their smaller physical size and easier fabrication and maintenance, injected diode laser systems appear to be easier to scale up than frequency-doubled systems. Additional requirements on the Raman lasers used for driving gates are a pure frequency spectrum, stable intensity and good phase coherence between the two Raman beams. Injected laser diodes inherently have a very stable intensity and good phase coherence.

\begin{figure}[bh!]
\centering
\includegraphics[width=\linewidth]{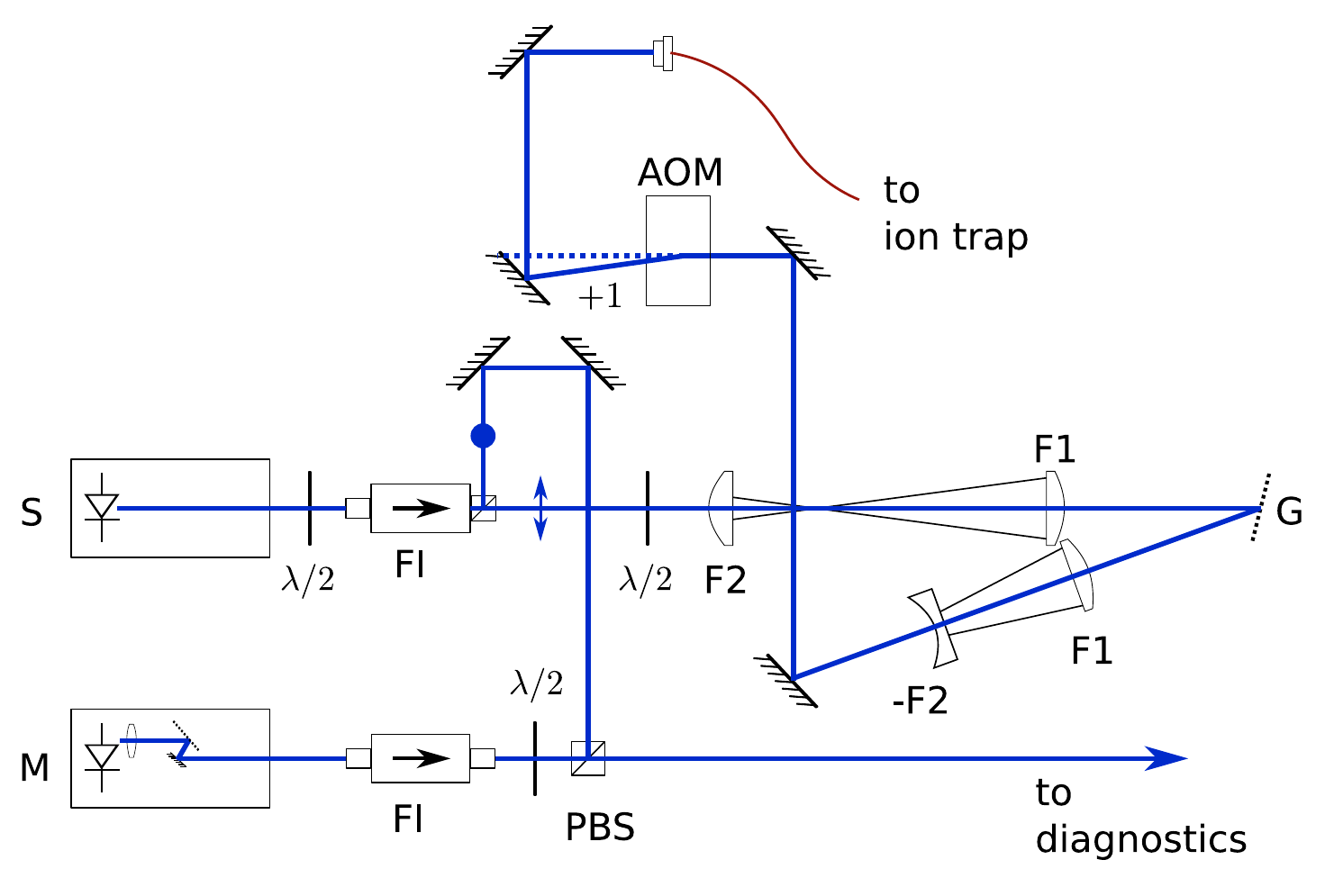}
\caption{(Color online) Experimental setup. The master laser (M) has a grating in Littrow configuration that allows for frequency tuning. The injection power can be adjusted with the $\lambda/2$ plate in front of the polarizing beam splitter (PBS). For frequency stabilisation the master laser is locked with a side-of-fringe lock to a low-drift cavity with finesse $\mathcal{F}\approx15$. The master and slave laser can both be monitored on a wave meter and an optical spectrum analyser. The slave laser (S) beam is diffracted by a grating (G) to filter out ASE (amplified spontaneous emission) light. A second slave laser was added to the system and injected by the master laser, with a beam path equivalent to that of the first slave laser. FI: Faraday isolator; AOM: acousto-optic modulator; F1,F2: lenses.}
\label{fig:setup}
\end{figure}

In this paper we demonstrate injection-locking of a pair of nominal $120\mW$ high-power laser diodes. The spectral purity of the injected lasers is investigated with scattering rate experiments on \cf and improved by inserting a diffraction grating and telescope into the beam path. The intensity stability is measured on a photodiode and compared to that of a frequency-doubled system. The phase coherence is investigated via an optical heterodyne experiment. 

\begin{figure}[th!]
\centering
\includegraphics[width=\linewidth]{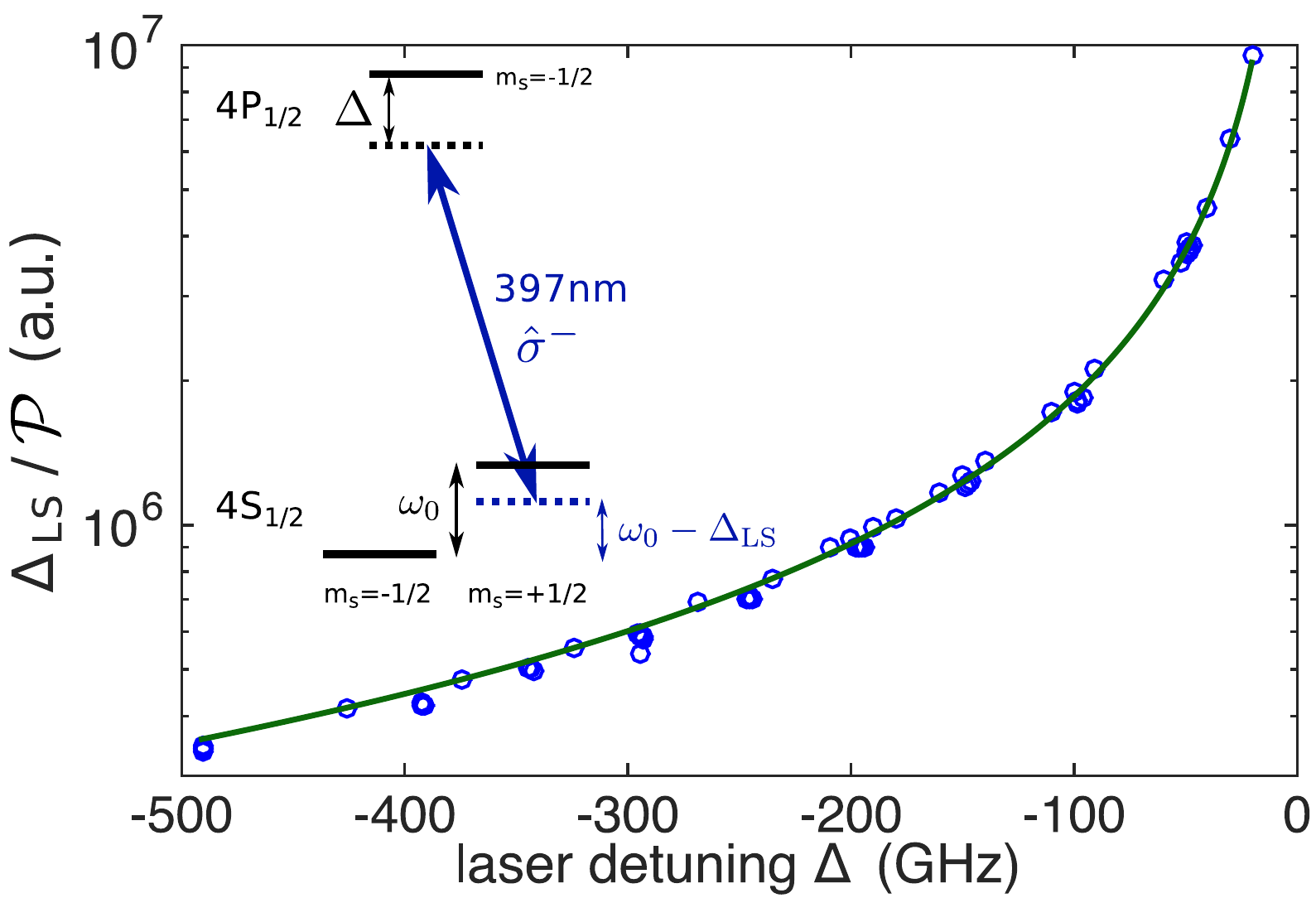}
\caption{(Color online) Ratio of light shift $\Delta_\mathrm{LS}$ to laser power $\mathcal{P}$. The green line corresponds to the theoretical light shift with fitted coupling strength $g$. The light shift in contrast to the scattering rate shows no visible ASE peaks. The reason for this is that the power in each ASE peak, as deduced from the scattering rate measurements, is only about $\mathcal{P}_\mathrm{ASE}/\mathcal{P}_\mathrm{carrier}\approx 3\times 10^{-4}$ of the carrier power, and the form of the ASE distribution leads to a high degree of cancellation in the light shift. The inset shows the \cf levels involved in the light shift measurements.}
\label{fig:Inj_ls}
\end{figure}

The injection setup is shown in Fig.\ref{fig:setup}. Light from the master laser is injected into the slave diode via a side port of the Faraday isolator. The output of the slave laser is directed onto a grating and the 1st order diffracted beam then goes through an AOM. The first-order diffracted beam of the AOM then goes through a single-mode fibre to the trapped ion.
The slave laser diodes are $397\pm0.5\nm$ frequency-selected laser diodes from Nichia (NDU4316E) mounted in a Toptica DL100 laser head. The master laser is a $30\mW$ grating-stabilised extended cavity diode laser.
For setting up the injection, the spatial modes are matched via their back-reflections with the slave diode current slightly below threshold. Injection at higher currents is achieved by adjusting the slave temperature, the slave current and the injection beam power for a given master laser frequency. Successful injection was demonstrated over a wide range of frequencies ($\pm 800\GHz$ or $\pm 0.4\nm$).

The slave temperature was usually kept at $T_s=22.1\dg$, but for large detunings $\Delta$, where the mode matching between the master laser and the slave diode is worse, other temperatures can be favourable. For $\Delta=435\GHz$ stable injection was achieved for temperatures $15\dg<T_s<25\dg$.
When increasing the slave current, modes of stable injection appear and disappear. The modes are $\approx 1-2\mA$ wide and spaced by $\approx 13\mA$. The position of the modes depends on $\Delta$ and $T_s$ and is subject to current hysteresis. The width of the modes decreases with increasing slave current. The slave current is limited by the damage threshold of the slave diode and the highest mode in these experiments was typically at $I_s\approx 90\mA$, giving $\mathcal{P}_s\approx 110\mW$.
The injection beam power has a large effect on the quality of injection. Typical injection powers are between $70\uW$ and $2\mW$. The necessary injection power strongly depends on the master frequency in a non-monotonic fashion. 
The fibre-coupling efficiency without the use of anamorphic prisms was typically $>50\%$.

Injection would typically stay stable over the course of a day. However the master laser needed occasional current adjustments to suppress parasitic frequency modes. For large detunings $|\Delta| \gtrsim 750\,\GHz$ the regions of stable injection of the slave laser would become smaller due to the worse mode matching. Due to the large current hysteresis of these modes the injection was therefore often lost within several minutes for higher current modes. However in the $1500\,\GHz$ wide region of frequencies that we were generally operating in, this did not occur.

\begin{figure}[th!]
\centering
\includegraphics[width=\linewidth]{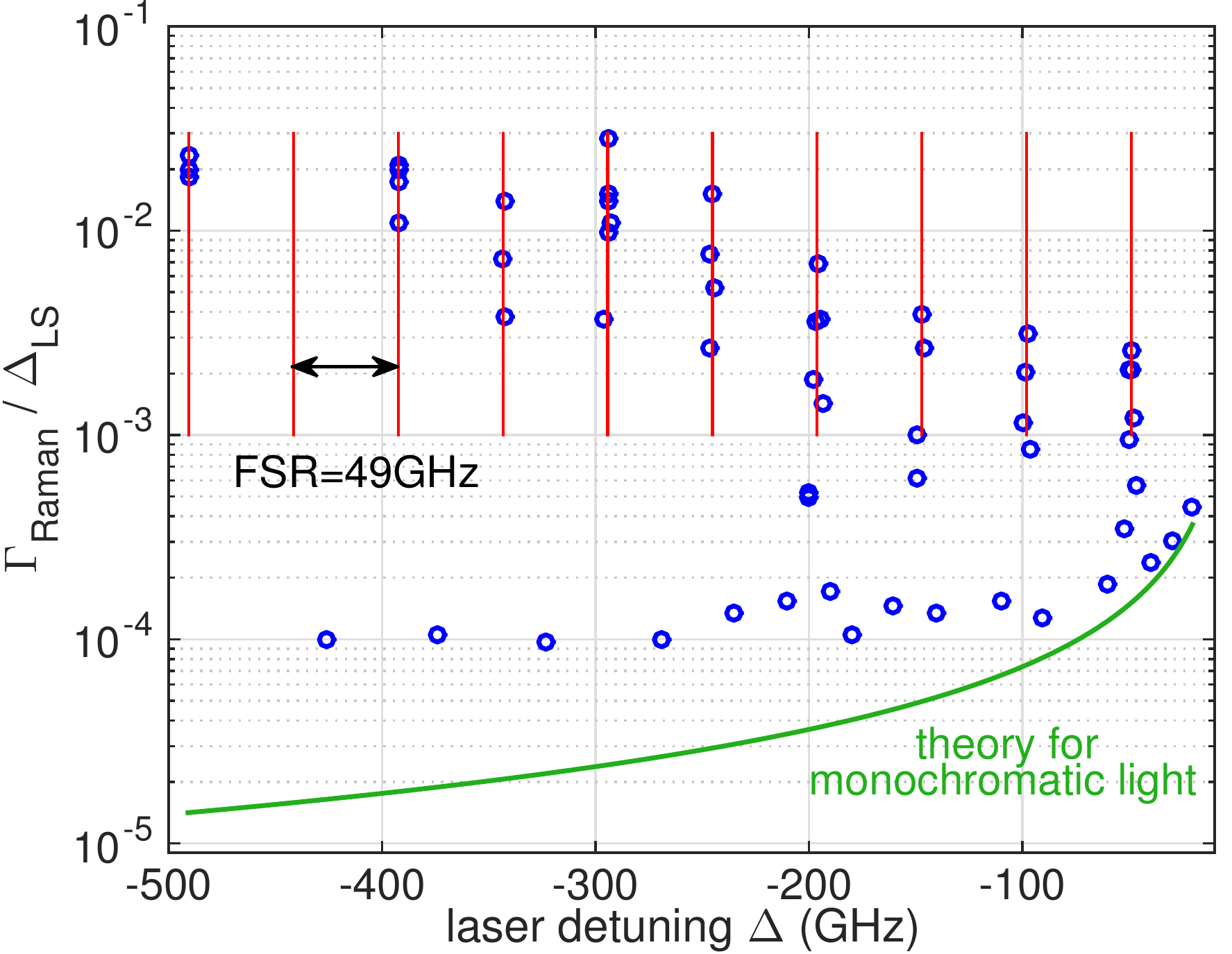}
\caption{(Color online) Ratio of the scattering rate $\Gamma_\mathrm{Raman}$ and the differential light shift $\Delta_\mathrm{LS}$ of the injected slave laser on \cf without grating or telescope inserted. $\Gamma_\mathrm{Raman}$ was measured by preparing the ion in the $4\mathrm{S}_{1/2}^{+1/2}$ state, and measuring the optical pumping rate to the $4\mathrm{S}_{1/2}^{-1/2}$ state when a far-detuned $397\nm$ $\hat{\sigma}_-$-polarized beam is applied. The scattering rate lies clearly above the theoretical line for monochromatic light, even when the detuning is not close to an amplified spontaneous emission ASE peak (red vertical lines). Additionally there are strong ASE peaks visible, which are spaced by the free spectral range FSR=49GHz of the internal cavity of the slave laser and are approximately FWHM=2GHz broad. Statistical errors are smaller than the size of the markers.}
\label{fig:Inj_scat}
\end{figure}

The spectral purity of the injected laser is investigated by measuring the Raman scattering rate. 
For this the ion was prepared in $\up=4S_{1/2}^{+1/2}$ by optical pumping with a $\hat{\sigma}_+$-polarized 397nm laser. Subsequently the $\hat{\sigma}_-$-polarized far-detuned (by $\Delta$) 397nm injected laser was applied for a time $t_\mathrm{scat}$. During this time the ion can be off-resonantly excited to the $4P_{1/2}$ or $4P_{3/2}$ levels, and can either relax back into \up by elastic (Rayleigh) scattering, or to $\dn=4S_{1/2}^{-1/2}$ or the $3D$ manifold by inelastic (Raman) scattering. Next, population in \dn is transferred state-selectively to the long lived $3D_{5/2}$ levels \cite{ReadMCD}. Fluorescence detection is used to determine whether the ion is still in the $4S_{1/2}$ state, allowing us to infer the population originally in \dn. By fitting the exponential decrease in population in \up as a function of pulse length $t_\mathrm{scat}$, the scattering rate can be deduced. 
For large detunings $|\Delta|\gg\gamma$, the Raman scattering rate \cite{HypOLJ} can be described as $\Gamma_{\downarrow\leftarrow\uparrow,\mathrm{Raman}}=\frac{\gamma g^2 (e_0^2+e_-^2)}{18}\left(\frac{\omega_f}{\Delta(\Delta-\omega_f)}\right)^2$, with $\gamma$ the natural linewidth of the excited state levels ($4P_{1/2}$ and $4P_{3/2}$), polarization vector $\bm{\varepsilon} =e_0 \hat{\pi} + e_+ \hat{\sigma}_+ + e_- \hat{\sigma}_-$, $g$ the coupling strength to the lasers and $\omega_f$ the fine structure splitting between the $4P_{1/2}$ and $4P_{3/2}$ manifolds. 
For $\Delta\rightarrow 0$ the scattering rate increases strongly and is therefore a good measure for near-resonant parasitic frequencies in primarily monochromatic light. Variations in intensity are normalized by taking the ratio with the differential light shift \cite{Win2003} $\Gamma_{\downarrow\leftarrow\uparrow,\mathrm{Raman}} /\Delta_{LS}=\frac{e_0^2+e_-^2}{e_+^2-e_-^2}\frac{\gamma\omega_f}{3\Delta(\Delta-\omega_f)}$. Even though the light shift also has a frequency dependence $\Delta_{LS}\propto\frac{1}{\Delta}$, the spectral impurities are too small to cause deviations from the light shift caused by monochromatic light, see Fig. \ref{fig:Inj_ls}. The light shift is therefore suitable to be used for intensity normalization.
We measured $\Delta_\mathrm{LS}$ by Ramsey spectroscopy of the $4\mathrm{S}_{1/2}^{-1/2}- {4\mathrm{S}_{1/2}^{+1/2}}$ splitting in a spin-echo sequence \cite{SeHAHN}, with the far-detuned $\hat{\sigma}_-$-polarized $397 \nm$ beam applied in the second half for varying durations $t_\mathrm{LS}$. The pulse sequence is $\mathrm{R}(\pi/2) - \mathrm{wait}[\tau] - \mathrm{R}(\pi) - 397_\mathrm{LS}[t_\mathrm{LS}] - \mathrm{wait}[\tau-t_\mathrm{LS}] - \mathrm{R}(\pi/2)$, where $R(\theta)$ are spin-rotations by an angle $\theta$ and are carried out by RF-radiation. State preparation and readout methods are the same as in the scattering rate measurements.
Fig. \ref{fig:Inj_scat} shows the measured scattering rate with scanned detuning $\Delta$ from resonance. There are strong peaks in the scattering rate visible, spaced $49\GHz$ apart from each other. This spacing corresponds to the free spectral range of the  internal cavity of the slave laser. The peaks are caused by amplified spontaneous emission (ASE) of the intrinsic modes of the slave cavity that are amplified by the gain medium but stay below lasing threshold \cite{CDn}. The wings of these peaks cause the scattering rate to be larger than that of monochromatic light for all frequencies, and to stay approximately constant for large detunings rather than to decay strongly. This would therefore increase the scattering errors and is unsuitable for high fidelity gate operations.

\begin{figure}[bh!]
\centering
\includegraphics[width=\linewidth]{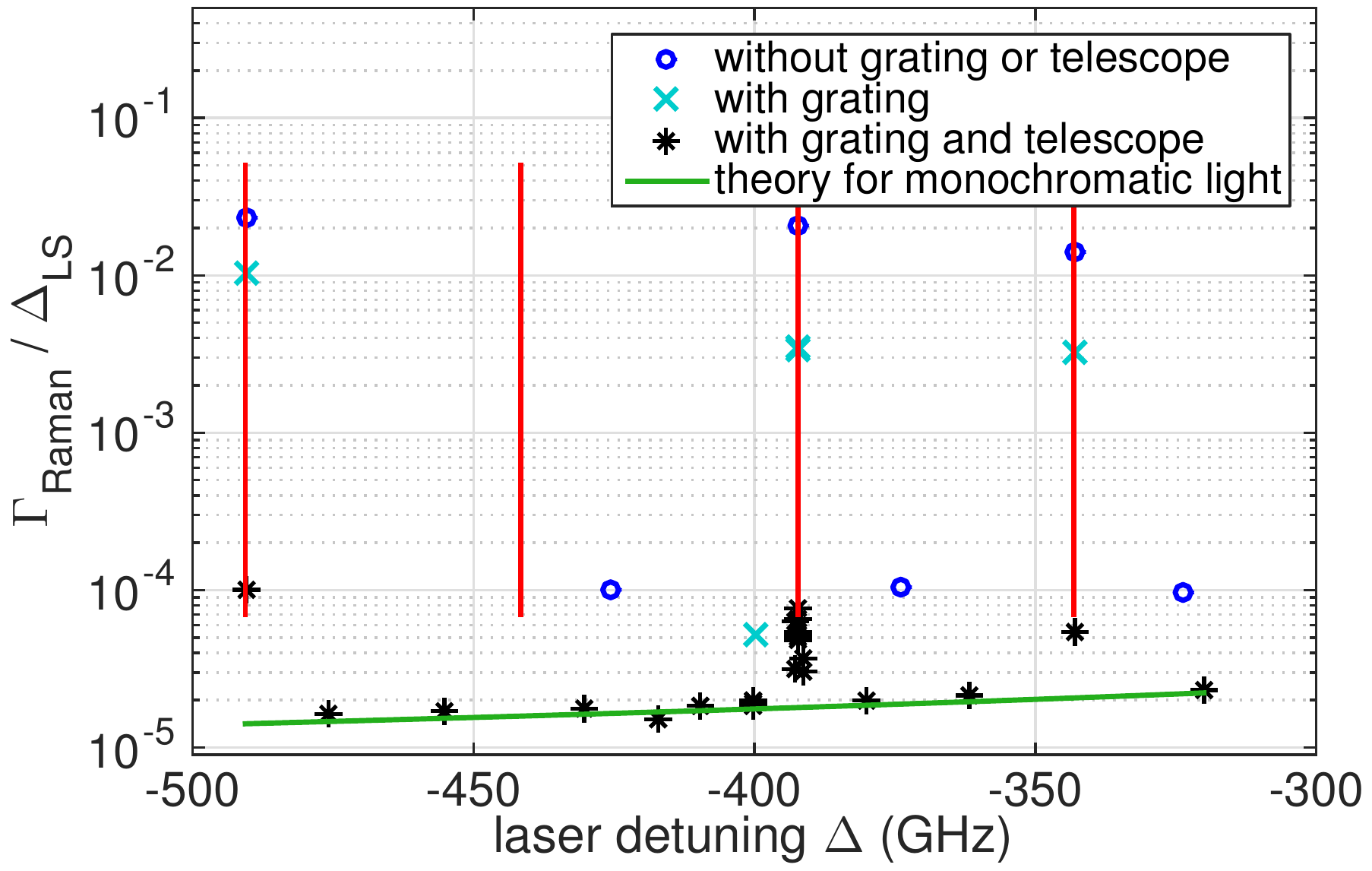}\\
\caption{(Color online) Comparison of the scattering rates with a grating and a telescope inserted into the beam path. Both the ASE peaks (red lines) and the baseline scattering rate are strongly decreased. Between the ASE peaks, agreement with the theoretical curve suggests that the extinction is good enough that the scattering rate is dominated by the scattering rate of monochromatic light and not by spectral impurities of the slave laser. Statistical errors are smaller than the size of the markers.}
\label{fig:Inj_scat_grat}
\end{figure}

\begin{figure}[bh!]
\raggedright
(a)
\includegraphics[width=\linewidth]{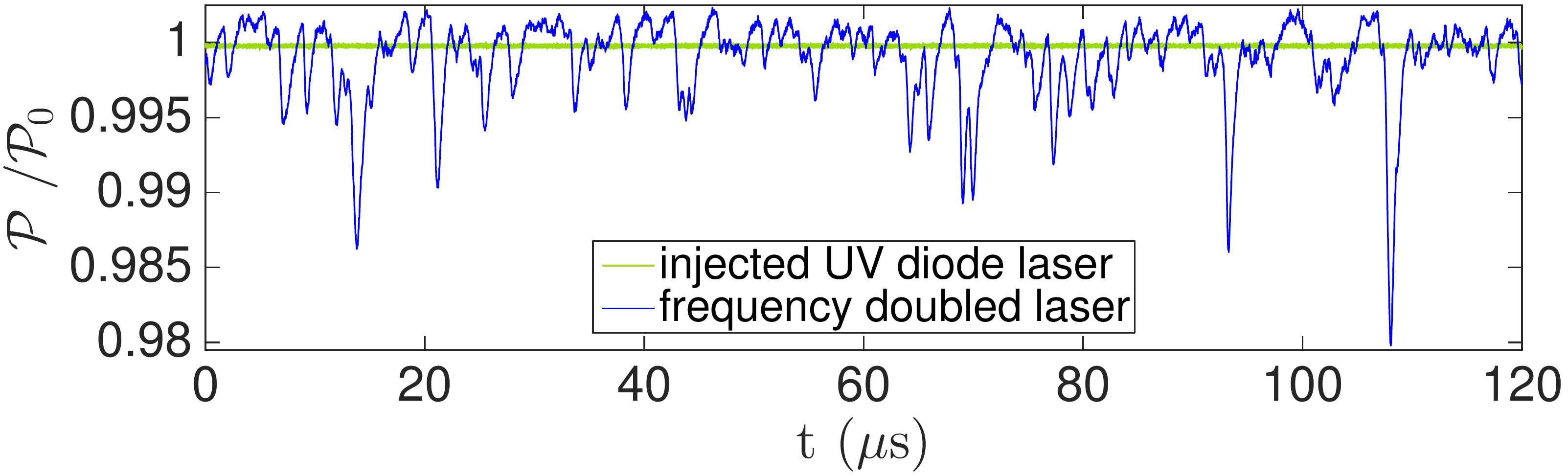} \\
(b)
\includegraphics[width=\linewidth]{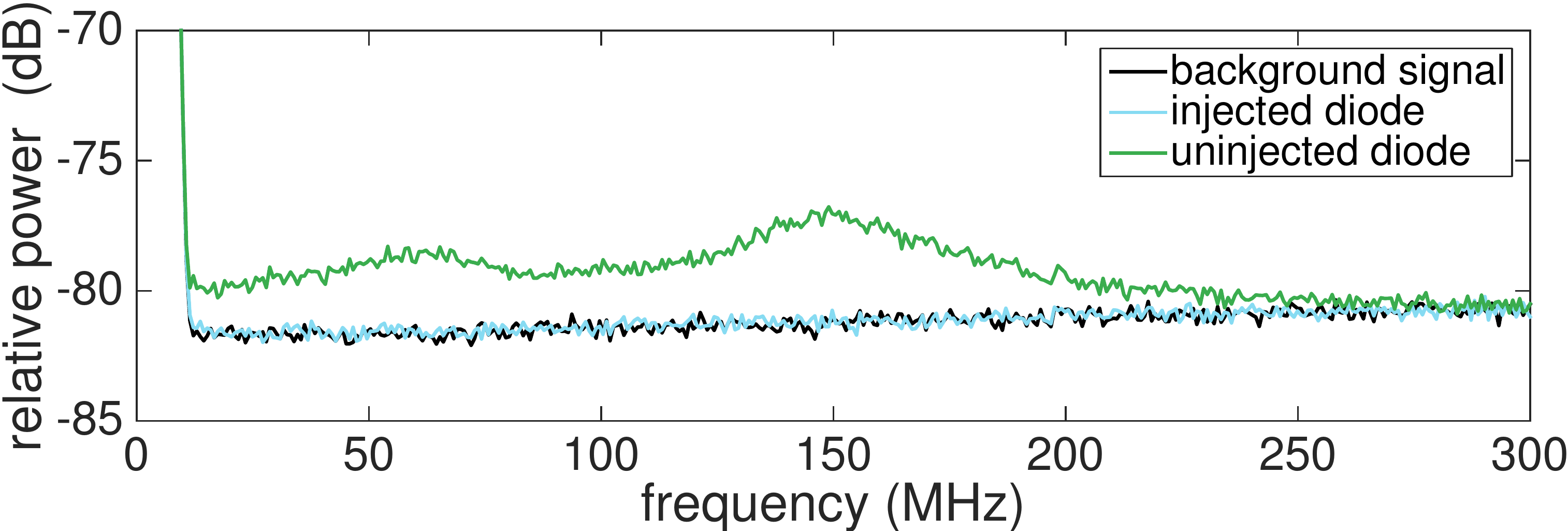} 
\caption{(Color online) {\bf(a)} Measurements of the noise of the normalized optical power in the time domain. A commercially available laser system with a frequency-doubling cavity \citep{InjNML} has fast intensity fluctuations at the 2\% level, whereas the noise of the injected diode laser is indistinguishable from the photodiode noise. {\bf(b)} Power noise measurements in the the frequency domain. There is a clear drop in the noise of the slave laser after it is injection locked, compared to when it is lasing freely without injection, which can be used as a sensitive diagnostic of the injection quality. Again the noise of the injected diode laser is indistinguishable from the photodiode noise.}
\label{fig:Inj_inoise}
\end{figure}

To improve the spectral purity, a blazed grating (Thorlabs GR25-1205) was inserted into the slave beam. This changes the beam direction of parasitic frequencies compared to the desired frequency, and therefore also their beam position at the $\approx1\m$ distant AOM and fibre. Thus the diffraction efficiency and coupling efficiency of parasitic frequencies decreases. The displacement for frequencies detuned by $\Delta = 400\GHz$ from the carrier differs by $\delta x=0.5\mm$ from the displacement of the carrier light at a distance of $1\m$ from the grating. The FWHM of the diffraction maximum for our beam with $w_0=1.2\mm$ at 1m distance is $0.43\mm$. To further increase the resolving power of the grating a telescope was inserted into the beam. This increases the spot size on the grating by a factor of 2.5, reducing the FWHM to $0.17\mm$.
Fig. \ref{fig:Inj_scat_grat} shows the improved scattering data. With both grating and telescope inserted parasitic modes are suppressed by $10^2-10^{3}$ and the scattering baseline corresponds to the scattering rate of monochromatic light. The grating and telescope reduce the power available at the fibre output by $43\%$.

The spectral purity of frequency-doubled systems is inherently very good, since the doubling cavity acts as a frequency filter. Mechanical vibrations of the cavity can however lead to intensity fluctuations. Injected diode lasers on the other hand have an inherently very stable intensity. Fig.\,\ref{fig:Inj_inoise} (a) shows the intensity stability of the injected slave laser measured with a $1\GHz$ bandwidth photodiode (Hamamatsu S5973) on an oscilloscope compared to a frequency doubled system. 
The noise of the injected slave laser light is indistinguishable from the photodiode noise floor, equivalent to an optical power of $\mathcal{P}_\mathrm{noise}=1.4\uW$ and therefore $\frac{\mathcal{P}_\mathrm{noise}}{\mathcal{P}}<2.2\times 10^{-4}$, which corresponds to an improvement of at least two orders of magnitude compared to the frequency-doubled system.
Analysis of the photodiode signal with a spectrum analyser shows a clear drop in intensity noise of the injected beam compared to a poorly injected or uninjected beam, see Fig.\,\ref{fig:Inj_inoise}(b).

\begin{figure}[th!]
\raggedright
(a)
\includegraphics[width=\linewidth]{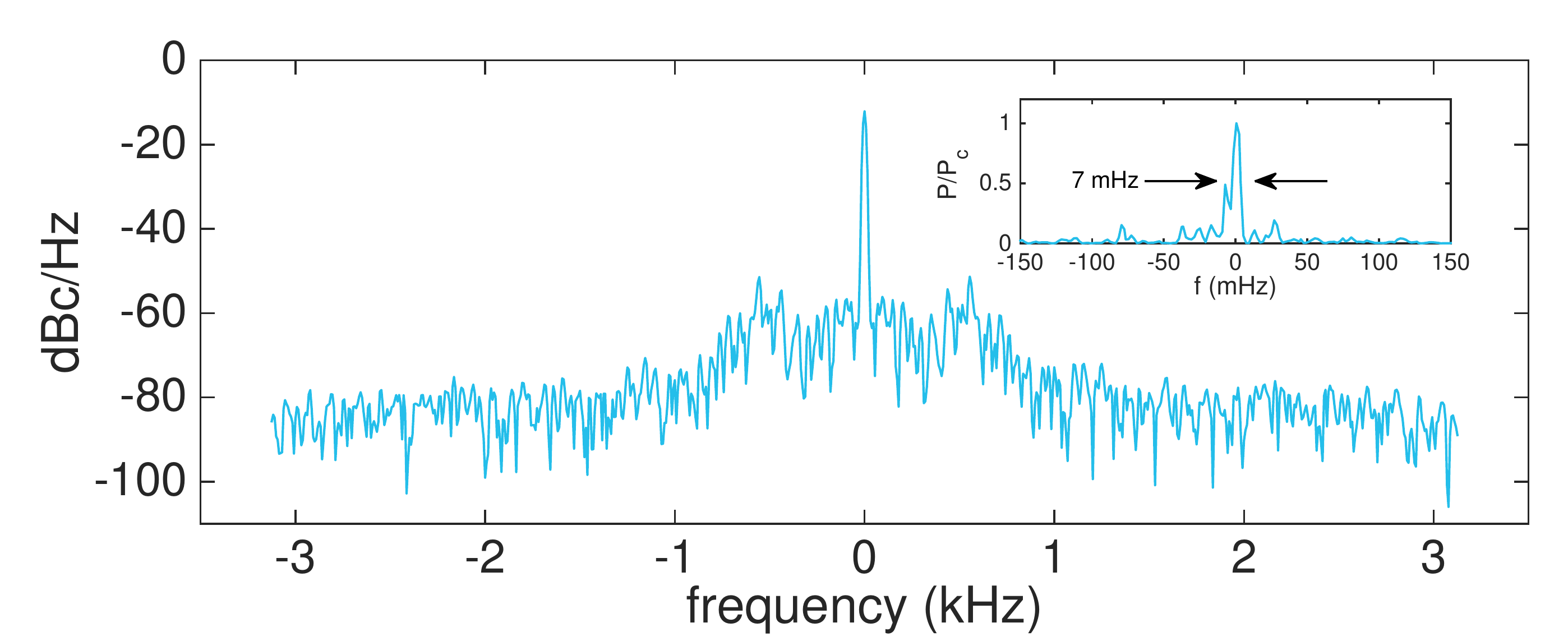} \\
(b)
\includegraphics[width=\linewidth]{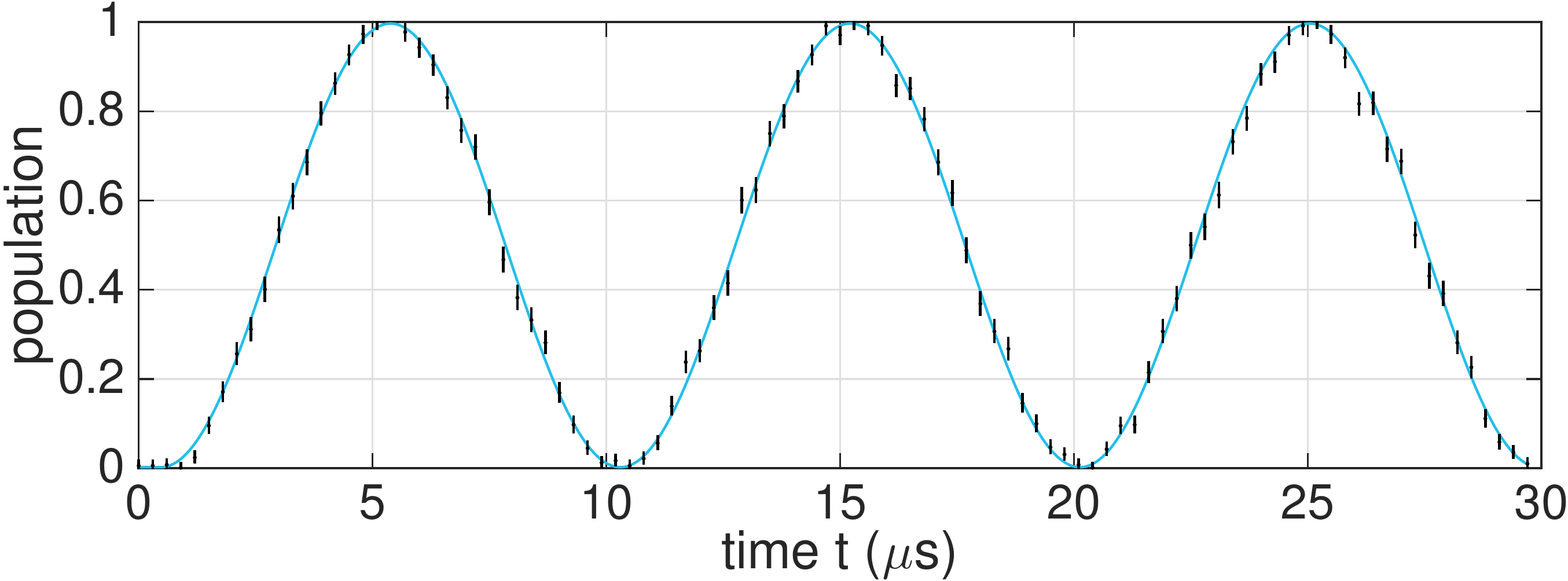}
\caption{(Color online) {\bf (a)} Beat note spectrum of the two injected slave lasers after subtracting a 10kHz frequency offset. The power is mainly in the central peak with a weak acoustic pedestal which is likely caused by vibrations of elements in the beam path and air currents. The inset shows a higher resolution scan of the center of the peak, plotted on a linear scale. The interaction linewidth of the Raman lasers extracted from the beat note is 7mHz, probably limited by measurement resolution.  {\bf (b)} Rabi flopping driven by the two injected slave lasers with $\Delta=400\,\GHz$ and $\mathcal{P}=4\mW$ per beam at the ion. State preparation and readout errors are normalized out. The blue line is a fit to the data with Rabi frequency $\Omega_R=2\pi\cdot 102\, \kHz$. The time offset $t=0.5\us$ at the beginning is caused by the difference in beam switching delay between the two AOMs due to different propagation distances of the acoustic waves in the AOM crystal from the transducer to the laser beam. }
\label{fig:Rabi}
\end{figure}

For driving Raman transitions two phase-coherent lasers are necessary, with the transition frequency as their frequency difference. Therefore a second slave laser was added to the setup, which was also injected by the master laser. The phase coherence of the two injected lasers was measured with a heterodyne beat-note experiment with a 10kHz frequency splitting of the two lasers induced by the AOMs in the beampath. The laser beams were combined on a non-polarizing beam-splitter and focussed onto an amplified photodiode about 2m distant from the laser output, after the telescope and the AOM. Both beams had approximately equal power and the same polarization. The signal from the photodiode was analysed on a Fourier-transform spectrum analyser and the result can be seen in Fig.\,\ref{fig:Rabi}\,(a).
The two injected lasers were also used to drive Rabi oscillations on the {\up}-\dn transition, see Fig.\,\ref{fig:Rabi}\,(b). The necessary frequency difference of $5\MHz$ was set by the two AOMs.

In this paper we presented a pair of injected high-power UV diode lasers with intensity stability $\lesssim 0.02\%$ and spectral purity indistinguishable from that of a monochromatic source over most of the gain profile. The lasers stay stably injected over a day and require injection powers as low as $70\uW$.

Recently injection of a high-power UV 399nm diode laser has also been reported \cite{InjKoz}.\\

\section*{Funding Information}
This work was supported by the UK Engineering and Physical Sciences Research Council (EPSRC).

\section*{Acknowledgements}

We thank Derek Stacey for helpful discussions and useful comments on the manuscript.

\bibliography{bibSub}

\begin{thebibliography}{16}%
\makeatletter
\providecommand \@ifxundefined [1]{%
 \@ifx{#1\undefined}
}%
\providecommand \@ifnum [1]{%
 \ifnum #1\expandafter \@firstoftwo
 \else \expandafter \@secondoftwo
 \fi
}%
\providecommand \@ifx [1]{%
 \ifx #1\expandafter \@firstoftwo
 \else \expandafter \@secondoftwo
 \fi
}%
\providecommand \natexlab [1]{#1}%
\providecommand \enquote  [1]{``#1''}%
\providecommand \bibnamefont  [1]{#1}%
\providecommand \bibfnamefont [1]{#1}%
\providecommand \citenamefont [1]{#1}%
\providecommand \href@noop [0]{\@secondoftwo}%
\providecommand \href [0]{\begingroup \@sanitize@url \@href}%
\providecommand \@href[1]{\@@startlink{#1}\@@href}%
\providecommand \@@href[1]{\endgroup#1\@@endlink}%
\providecommand \@sanitize@url [0]{\catcode `\\12\catcode `\$12\catcode
  `\&12\catcode `\#12\catcode `\^12\catcode `\_12\catcode `\%12\relax}%
\providecommand \@@startlink[1]{}%
\providecommand \@@endlink[0]{}%
\providecommand \url  [0]{\begingroup\@sanitize@url \@url }%
\providecommand \@url [1]{\endgroup\@href {#1}{\urlprefix }}%
\providecommand \urlprefix  [0]{URL }%
\providecommand \Eprint [0]{\href }%
\providecommand \doibase [0]{http://dx.doi.org/}%
\providecommand \selectlanguage [0]{\@gobble}%
\providecommand \bibinfo  [0]{\@secondoftwo}%
\providecommand \bibfield  [0]{\@secondoftwo}%
\providecommand \translation [1]{[#1]}%
\providecommand \BibitemOpen [0]{}%
\providecommand \bibitemStop [0]{}%
\providecommand \bibitemNoStop [0]{.\EOS\space}%
\providecommand \EOS [0]{\spacefactor3000\relax}%
\providecommand \BibitemShut  [1]{\csname bibitem#1\endcsname}%
\let\auto@bib@innerbib\@empty
\bibitem [{\citenamefont {Wineland}\ \emph {et~al.}(1978)\citenamefont
  {Wineland}, \citenamefont {Drullinger},\ and\ \citenamefont
  {Walls}}]{DyeWin}%
  \BibitemOpen
  \bibfield  {author} {\bibinfo {author} {\bibfnamefont {D.~J.}\ \bibnamefont
  {Wineland}}, \bibinfo {author} {\bibfnamefont {R.~E.}\ \bibnamefont
  {Drullinger}}, \ and\ \bibinfo {author} {\bibfnamefont {F.~L.}\ \bibnamefont
  {Walls}},\ }\href@noop {} {\bibfield  {journal} {\bibinfo  {journal} {Phys.
  Rev. Lett.}\ }\textbf {\bibinfo {volume} {40}} (\bibinfo {year}
  {1978})}\BibitemShut {NoStop}%
\bibitem [{\citenamefont {Wilson}\ \emph {et~al.}(2011)\citenamefont {Wilson},
  \citenamefont {Ospelklaus}, \citenamefont {VanDevender}, \citenamefont
  {Mlynek}, \citenamefont {Leibfried},\ and\ \citenamefont
  {Wineland}}]{LasWIL}%
  \BibitemOpen
  \bibfield  {author} {\bibinfo {author} {\bibfnamefont {A.}~\bibnamefont
  {Wilson}}, \bibinfo {author} {\bibfnamefont {C.}~\bibnamefont {Ospelklaus}},
  \bibinfo {author} {\bibfnamefont {A.}~\bibnamefont {VanDevender}}, \bibinfo
  {author} {\bibfnamefont {J.}~\bibnamefont {Mlynek}}, \bibinfo {author}
  {\bibfnamefont {D.}~\bibnamefont {Leibfried}}, \ and\ \bibinfo {author}
  {\bibfnamefont {D.}~\bibnamefont {Wineland}},\ }\href@noop {} {\bibfield
  {journal} {\bibinfo  {journal} {Appl. Phys. B}\ }\textbf {\bibinfo {volume}
  {105}},\ \bibinfo {pages} {741} (\bibinfo {year} {2011})}\BibitemShut
  {NoStop}%
\bibitem [{\citenamefont {Lo}\ \emph {et~al.}(2013)\citenamefont {Lo},
  \citenamefont {Alonso}, \citenamefont {Kienzler}, \citenamefont {Keitch},
  \citenamefont {de~Clercq}, \citenamefont {Negnevitsky},\ and\ \citenamefont
  {Home}}]{LasHLO}%
  \BibitemOpen
  \bibfield  {author} {\bibinfo {author} {\bibfnamefont {H.-Y.}\ \bibnamefont
  {Lo}}, \bibinfo {author} {\bibfnamefont {J.}~\bibnamefont {Alonso}}, \bibinfo
  {author} {\bibfnamefont {D.}~\bibnamefont {Kienzler}}, \bibinfo {author}
  {\bibfnamefont {B.}~\bibnamefont {Keitch}}, \bibinfo {author} {\bibfnamefont
  {L.}~\bibnamefont {de~Clercq}}, \bibinfo {author} {\bibfnamefont
  {V.}~\bibnamefont {Negnevitsky}}, \ and\ \bibinfo {author} {\bibfnamefont
  {J.}~\bibnamefont {Home}},\ }\href@noop {} {\bibfield  {journal} {\bibinfo
  {journal} {Appl. Phys. B}\ }\textbf {\bibinfo {volume} {114}},\ \bibinfo
  {pages} {17} (\bibinfo {year} {2013})}\BibitemShut {NoStop}%
\bibitem [{\citenamefont {Linke}\ \emph {et~al.}(2013)\citenamefont {Linke},
  \citenamefont {Ballance},\ and\ \citenamefont {Lucas}}]{InjNML}%
  \BibitemOpen
  \bibfield  {author} {\bibinfo {author} {\bibfnamefont {N.}~\bibnamefont
  {Linke}}, \bibinfo {author} {\bibfnamefont {C.}~\bibnamefont {Ballance}}, \
  and\ \bibinfo {author} {\bibfnamefont {D.}~\bibnamefont {Lucas}},\
  }\href@noop {} {\bibfield  {journal} {\bibinfo  {journal} {Optics Letters}\
  }\textbf {\bibinfo {volume} {38}},\ \bibinfo {pages} {5087} (\bibinfo {year}
  {2013})}\BibitemShut {NoStop}%
\bibitem [{\citenamefont {Diddams}\ \emph {et~al.}(2004)\citenamefont
  {Diddams}, \citenamefont {Bergquist}, \citenamefont {Jefferts},\ and\
  \citenamefont {Oates}}]{ClkDid}%
  \BibitemOpen
  \bibfield  {author} {\bibinfo {author} {\bibfnamefont {S.}~\bibnamefont
  {Diddams}}, \bibinfo {author} {\bibfnamefont {J.}~\bibnamefont {Bergquist}},
  \bibinfo {author} {\bibfnamefont {S.}~\bibnamefont {Jefferts}}, \ and\
  \bibinfo {author} {\bibfnamefont {C.}~\bibnamefont {Oates}},\ }\href@noop {}
  {\bibfield  {journal} {\bibinfo  {journal} {Science}\ }\textbf {\bibinfo
  {volume} {306}},\ \bibinfo {pages} {1318} (\bibinfo {year}
  {2004})}\BibitemShut {NoStop}%
\bibitem [{\citenamefont {Schmidt}\ \emph {et~al.}(2005)\citenamefont
  {Schmidt}, \citenamefont {Rosenband}, \citenamefont {Langer}, \citenamefont
  {Itano}, \citenamefont {Bergquist},\ and\ \citenamefont
  {Wineland}}]{SpecSCHMIDT}%
  \BibitemOpen
  \bibfield  {author} {\bibinfo {author} {\bibfnamefont {P.}~\bibnamefont
  {Schmidt}}, \bibinfo {author} {\bibfnamefont {T.}~\bibnamefont {Rosenband}},
  \bibinfo {author} {\bibfnamefont {C.}~\bibnamefont {Langer}}, \bibinfo
  {author} {\bibfnamefont {W.}~\bibnamefont {Itano}}, \bibinfo {author}
  {\bibfnamefont {J.}~\bibnamefont {Bergquist}}, \ and\ \bibinfo {author}
  {\bibfnamefont {D.}~\bibnamefont {Wineland}},\ }\href@noop {} {\bibfield
  {journal} {\bibinfo  {journal} {Science}\ }\textbf {\bibinfo {volume} {309}}
  (\bibinfo {year} {2005})}\BibitemShut {NoStop}%
\bibitem [{\citenamefont {Roos}\ \emph {et~al.}(2006)\citenamefont {Roos},
  \citenamefont {Chwalla}, \citenamefont {Kim}, \citenamefont {Riebe},\ and\
  \citenamefont {Blatt}}]{SpecROOS}%
  \BibitemOpen
  \bibfield  {author} {\bibinfo {author} {\bibfnamefont {C.}~\bibnamefont
  {Roos}}, \bibinfo {author} {\bibfnamefont {M.}~\bibnamefont {Chwalla}},
  \bibinfo {author} {\bibfnamefont {K.}~\bibnamefont {Kim}}, \bibinfo {author}
  {\bibfnamefont {M.}~\bibnamefont {Riebe}}, \ and\ \bibinfo {author}
  {\bibfnamefont {R.}~\bibnamefont {Blatt}},\ }\href@noop {} {\bibfield
  {journal} {\bibinfo  {journal} {Nature}\ }\textbf {\bibinfo {volume} {443}}
  (\bibinfo {year} {2006})}\BibitemShut {NoStop}%
\bibitem [{\citenamefont {Monroe}\ and\ \citenamefont {Kim}(2013)}]{ScalMON}%
  \BibitemOpen
  \bibfield  {author} {\bibinfo {author} {\bibfnamefont {C.}~\bibnamefont
  {Monroe}}\ and\ \bibinfo {author} {\bibfnamefont {J.}~\bibnamefont {Kim}},\
  }\href@noop {} {\bibfield  {journal} {\bibinfo  {journal} {Science}\ }\textbf
  {\bibinfo {volume} {8}},\ \bibinfo {pages} {1164} (\bibinfo {year}
  {2013})}\BibitemShut {NoStop}%
\bibitem [{\citenamefont {Ballance}\ \emph {et~al.}(2014)\citenamefont
  {Ballance}, \citenamefont {Harty}, \citenamefont {Linke},\ and\ \citenamefont
  {Lucas}}]{HftqCJB}%
  \BibitemOpen
  \bibfield  {author} {\bibinfo {author} {\bibfnamefont {C.}~\bibnamefont
  {Ballance}}, \bibinfo {author} {\bibfnamefont {T.}~\bibnamefont {Harty}},
  \bibinfo {author} {\bibfnamefont {N.}~\bibnamefont {Linke}}, \ and\ \bibinfo
  {author} {\bibfnamefont {D.}~\bibnamefont {Lucas}},\ }\href@noop {}
  {\bibfield  {journal} {\bibinfo  {journal} {arXiv}\ }\textbf {\bibinfo
  {volume} {1406.5473}} (\bibinfo {year} {2014})}\BibitemShut {NoStop}%
\bibitem [{\citenamefont {Ozeri}\ \emph {et~al.}(2007)\citenamefont {Ozeri},
  \citenamefont {Itano}, \citenamefont {Blakestad}, \citenamefont {Britton},
  \citenamefont {Chiaverini}, \citenamefont {Jost}, \citenamefont {Langer},
  \citenamefont {Leibfried}, \citenamefont {Reichle}, \citenamefont {Seidelin},
  \citenamefont {Wesenberg},\ and\ \citenamefont {Wineland}}]{ErrOIB}%
  \BibitemOpen
  \bibfield  {author} {\bibinfo {author} {\bibfnamefont {R.}~\bibnamefont
  {Ozeri}}, \bibinfo {author} {\bibfnamefont {W.}~\bibnamefont {Itano}},
  \bibinfo {author} {\bibfnamefont {R.}~\bibnamefont {Blakestad}}, \bibinfo
  {author} {\bibfnamefont {J.}~\bibnamefont {Britton}}, \bibinfo {author}
  {\bibfnamefont {J.}~\bibnamefont {Chiaverini}}, \bibinfo {author}
  {\bibfnamefont {J.}~\bibnamefont {Jost}}, \bibinfo {author} {\bibfnamefont
  {C.}~\bibnamefont {Langer}}, \bibinfo {author} {\bibfnamefont
  {D.}~\bibnamefont {Leibfried}}, \bibinfo {author} {\bibfnamefont
  {R.}~\bibnamefont {Reichle}}, \bibinfo {author} {\bibfnamefont
  {S.}~\bibnamefont {Seidelin}}, \bibinfo {author} {\bibfnamefont
  {J.}~\bibnamefont {Wesenberg}}, \ and\ \bibinfo {author} {\bibfnamefont
  {D.}~\bibnamefont {Wineland}},\ }\href@noop {} {\bibfield  {journal}
  {\bibinfo  {journal} {Phys. Rev. A}\ }\textbf {\bibinfo {volume} {75}}
  (\bibinfo {year} {2007})}\BibitemShut {NoStop}%
\bibitem [{\citenamefont {McDonnell}\ \emph {et~al.}(2004)\citenamefont
  {McDonnell}, \citenamefont {Stacey}, \citenamefont {Webster}, \citenamefont
  {Home}, \citenamefont {Ramos}, \citenamefont {Lucas}, \citenamefont
  {D.N.Stacey},\ and\ \citenamefont {Steane}}]{ReadMCD}%
  \BibitemOpen
  \bibfield  {author} {\bibinfo {author} {\bibfnamefont {M.~J.}\ \bibnamefont
  {McDonnell}}, \bibinfo {author} {\bibfnamefont {J.}~\bibnamefont {Stacey}},
  \bibinfo {author} {\bibfnamefont {S.}~\bibnamefont {Webster}}, \bibinfo
  {author} {\bibfnamefont {J.}~\bibnamefont {Home}}, \bibinfo {author}
  {\bibfnamefont {A.}~\bibnamefont {Ramos}}, \bibinfo {author} {\bibfnamefont
  {D.}~\bibnamefont {Lucas}}, \bibinfo {author} {\bibnamefont {D.N.Stacey}}, \
  and\ \bibinfo {author} {\bibfnamefont {A.}~\bibnamefont {Steane}},\
  }\href@noop {} {\bibfield  {journal} {\bibinfo  {journal} {Phys. Rev. Lett.}\
  }\textbf {\bibinfo {volume} {93}},\ \bibinfo {pages} {153601} (\bibinfo
  {year} {2004})}\BibitemShut {NoStop}%
\bibitem [{\citenamefont {Ozeri}\ \emph {et~al.}(2005)\citenamefont {Ozeri},
  \citenamefont {Langer}, \citenamefont {Jost}, \citenamefont {DeMarco},
  \citenamefont {Ben-Kish}, \citenamefont {Blakestad}, \citenamefont {Britton},
  \citenamefont {Chiaverini}, \citenamefont {Itano}, \citenamefont {Hume},
  \citenamefont {Leibfried}, \citenamefont {Rosenband}, \citenamefont
  {Schmidt},\ and\ \citenamefont {Wineland}}]{HypOLJ}%
  \BibitemOpen
  \bibfield  {author} {\bibinfo {author} {\bibfnamefont {R.}~\bibnamefont
  {Ozeri}}, \bibinfo {author} {\bibfnamefont {C.}~\bibnamefont {Langer}},
  \bibinfo {author} {\bibfnamefont {J.}~\bibnamefont {Jost}}, \bibinfo {author}
  {\bibfnamefont {B.}~\bibnamefont {DeMarco}}, \bibinfo {author} {\bibfnamefont
  {A.}~\bibnamefont {Ben-Kish}}, \bibinfo {author} {\bibfnamefont
  {B.}~\bibnamefont {Blakestad}}, \bibinfo {author} {\bibfnamefont
  {J.}~\bibnamefont {Britton}}, \bibinfo {author} {\bibfnamefont
  {J.}~\bibnamefont {Chiaverini}}, \bibinfo {author} {\bibfnamefont
  {W.}~\bibnamefont {Itano}}, \bibinfo {author} {\bibfnamefont
  {D.}~\bibnamefont {Hume}}, \bibinfo {author} {\bibfnamefont {D.}~\bibnamefont
  {Leibfried}}, \bibinfo {author} {\bibfnamefont {T.}~\bibnamefont
  {Rosenband}}, \bibinfo {author} {\bibfnamefont {P.}~\bibnamefont {Schmidt}},
  \ and\ \bibinfo {author} {\bibfnamefont {D.}~\bibnamefont {Wineland}},\
  }\href@noop {} {\bibfield  {journal} {\bibinfo  {journal} {Phys. Rev. Lett.}\
  }\textbf {\bibinfo {volume} {95}} (\bibinfo {year} {2005})}\BibitemShut
  {NoStop}%
\bibitem [{\citenamefont {Wineland}\ \emph {et~al.}(2003)\citenamefont
  {Wineland}, \citenamefont {Barrett}, \citenamefont {Britton}, \citenamefont
  {Chiaverini}, \citenamefont {DeMarco}, \citenamefont {Itano}, \citenamefont
  {Jelenkovi\'{c}}, \citenamefont {Langer}, \citenamefont {Leibfried},
  \citenamefont {Meyer}, \citenamefont {Rosenband},\ and\ \citenamefont
  {Sch\"{a}tz}}]{Win2003}%
  \BibitemOpen
  \bibfield  {author} {\bibinfo {author} {\bibfnamefont {D.~J.}\ \bibnamefont
  {Wineland}}, \bibinfo {author} {\bibfnamefont {M.}~\bibnamefont {Barrett}},
  \bibinfo {author} {\bibfnamefont {J.}~\bibnamefont {Britton}}, \bibinfo
  {author} {\bibfnamefont {J.}~\bibnamefont {Chiaverini}}, \bibinfo {author}
  {\bibfnamefont {B.}~\bibnamefont {DeMarco}}, \bibinfo {author} {\bibfnamefont
  {W.~M.}\ \bibnamefont {Itano}}, \bibinfo {author} {\bibfnamefont
  {B.}~\bibnamefont {Jelenkovi\'{c}}}, \bibinfo {author} {\bibfnamefont
  {C.}~\bibnamefont {Langer}}, \bibinfo {author} {\bibfnamefont
  {D.}~\bibnamefont {Leibfried}}, \bibinfo {author} {\bibfnamefont
  {V.}~\bibnamefont {Meyer}}, \bibinfo {author} {\bibfnamefont
  {T.}~\bibnamefont {Rosenband}}, \ and\ \bibinfo {author} {\bibfnamefont
  {T.}~\bibnamefont {Sch\"{a}tz}},\ }\href@noop {} {\bibfield  {journal}
  {\bibinfo  {journal} {Philosophical transactions. Series A, Mathematical,
  physical, and engineering sciences}\ }\textbf {\bibinfo {volume} {361}},\
  \bibinfo {pages} {1349} (\bibinfo {year} {2003})}\BibitemShut {NoStop}%
\bibitem [{\citenamefont {Hahn}(1950)}]{SeHAHN}%
  \BibitemOpen
  \bibfield  {author} {\bibinfo {author} {\bibfnamefont {E.~L.}\ \bibnamefont
  {Hahn}},\ }\href@noop {} {\bibfield  {journal} {\bibinfo  {journal} {Phys.
  Rev.}\ }\textbf {\bibinfo {volume} {80}},\ \bibinfo {pages} {580} (\bibinfo
  {year} {1950})}\BibitemShut {NoStop}%
\bibitem [{\citenamefont {Donald}(2000)}]{CDn}%
  \BibitemOpen
  \bibfield  {author} {\bibinfo {author} {\bibfnamefont {C.}~\bibnamefont
  {Donald}},\ }\emph {\bibinfo {title} {Development of an Ion Trap Quantum
  Information Processor}},\ \href@noop {} {\bibinfo {type} {{D.P}hil.
  thesis}},\ \bibinfo  {school} {{U}niversity of {O}xford} (\bibinfo {year}
  {2000})\BibitemShut {NoStop}%
\bibitem [{\citenamefont {Hosoya}\ \emph {et~al.}(2014)\citenamefont {Hosoya},
  \citenamefont {Miranda}, \citenamefont {Inoue},\ and\ \citenamefont
  {Kozuma}}]{InjKoz}%
  \BibitemOpen
  \bibfield  {author} {\bibinfo {author} {\bibfnamefont {T.}~\bibnamefont
  {Hosoya}}, \bibinfo {author} {\bibfnamefont {M.}~\bibnamefont {Miranda}},
  \bibinfo {author} {\bibfnamefont {R.}~\bibnamefont {Inoue}}, \ and\ \bibinfo
  {author} {\bibfnamefont {M.}~\bibnamefont {Kozuma}},\ }\href@noop {}
  {\bibfield  {journal} {\bibinfo  {journal} {arXiv}\ }\textbf {\bibinfo
  {volume} {1412.0794v1}} (\bibinfo {year} {2014})}\BibitemShut {NoStop}%
\end{thebibliography}%

\end{document}